# Review of "Continuous Finite-Time Stabilization of Translational and Rotational Double Integrators"

Olalekan P. Ogunmolu

## Abstract

*We review the above-mentioned paper by [Bhat et. al., '98] where a class of bounded, continuous time-invariant finite time stabilizing feedback laws are derived for the double integrator and Lyapunov theory is employed in establishing finite-time convergence.*

## 1. Introduction

Feedback linearization often generates closed-loop Lipschitzian dynamics. In such systems, convergence is often exponential and carries the burden of decreased response time and repeated overshooting in the switching contour – as a result of the discontinuous control. This is undesirable in time-critical applications such as minimum-energy control, and conservation of momentum among others. Classical optimal control provides examples of systems that converge to the equilibrium in finite time such as the double integrator. Among approaches that have been employed to control the double integrator include open-loop methods and optimal synthesis.

Open-loop strategies involve minimizing a control energy cost function, $J(u)$, such that the state of the system is transferred from an initial conditions $x(0) = x_0$, $\dot{x}(0) = y_0$ to the equilibrium point in finite-time, $t_f$ [Athans et. al.]. An example of optimal synthesis is the bang-bang time-optimal controller where the minimization of a non-quadratic cost function subject to a saturation constraint on the control input yields a finite-time stabilizing feedback controller[Fuller, '66]. Time-optimal synthesis involve capturing the exact switching times for an open-loop control starting from the downward equilibrium.

Nonetheless, there are drawbacks in both cases with open-loop strategies being sensitive to system uncertainties and have poor disturbance rejection properties. Optimal synthesis methods are discontinuous with some of such controllers in [Fuller, '66] delivering cost functions that have infinite number of discontinuities such that they have little practical usefulness. It is therefore imperative to design continuous finite-time stabilizing controllers since they will have better robustness and disturbance rejection.

### Statement of the Problem

Consider a rigid body rotating under the action of a mechanical torque about a fixed axis. Its equations of motion resemble those of a double integrator. States differ by $2n\pi$ (where $n = 0, 1, 2, \ldots$) in angular modes which correspond to the same physical configuration of the body. State space for this system is $S^1 \times \mathbb{R}$ rather than $\mathbb{R}^2$ [Andronov et. al.]. Developing stabilizing controls for the double integrator on $\mathbb{R}^2$ (translational double integrator) will lead to unwinding since the configuration space is actually $\mathbb{R}$ This makes an interesting problem when designing feedback controllers for the rotational double integrator with anti-wind-up compensation. Discontinuous feedback controllers are practically infeasible due to the chattering they introduce because of plant uncertainties. They could also excite high-frequency dynamics when used in controlling lightly damped structures [Baruh et. al.]

## 2. Finite Time Systems

We will consider differential equations with an isolated equilibrium point at the origin, and no other equilibria for the sake of our study. A practical view would be the properties of a robot arm in the neighborhood of a set-point which is modeled as an isolated equilibrium and we are interested in the local behavior. If this origin is asymptotically stable, we will call solutions converging to the origin finite-time solutions. All right-hand sides of finite time differential equations will be $C^1$ everywhere except zero where they are assumed to be continuous.

It becomes apparent that the design of finite time stabilizing continuous time-invariant feedback controllers involve non-Lipschitzian closed-loop dynamics because as all solutions reach zero in finite time, there is non-uniqueness of solutions in backward time. Such non-unique (revert time) solutions would violate uniqueness conditions for Lipschitz differential equations.

## 2.1. Finite-Time Stabilization: A Definition

For the System of differential equations,

$$\dot{y}(t) = f(y(t)) \quad (1)$$

where $f : \mathscr{D} \mapsto \mathbb{R}^n$ is continuous on an open neighborhood $\mathscr{D} \subseteq \mathbb{R}^n$ of the origin and $f(0) = 0$, a continuously differentiable function $y : I \to \mathscr{D}$ is said to be a solution of (1) on the interval $I \subset \mathbb{R}$ if $y$ satisfies (1) for all $t \in I$.

We assume (??) possesses unique solutions in forward time except possibly at the origin for all initial conditions Uniqueness in forward time and the continuity of $f$ ensure that solutions are continuous functions of initial conditions even when $f$ is no longer Lipschitz continuous [Hartman et. al., '82, Th. 2.1, p. 94]

The origin is finite-time stable if there exists an open neighborhood $\mathscr{N} \subseteq \mathscr{D}$ of the origin and a settling time function $T : \mathscr{N} \setminus 0 \mapsto (0, \infty)$, such that we have the following:

1. Finite-time convergence: For every $x \in \mathscr{N} \setminus \{0\}$, $\rho_t(x)$ is defined for $t \in [0, T(x))$, $\rho_t(x) \in \mathscr{N} \setminus \{0\}$, for $t \in [0, T(x))$, and $\lim_{t \to T(x)} \rho_t(x) = 0$

2. Lyapunov stability: For every open set $\mathscr{U}_\varepsilon$ such that $0 \in \mathscr{U}_\varepsilon \subseteq \mathscr{N}$, there exists an open set $\mathscr{U}_\delta$ such that $0 \in \mathscr{U}_\delta \subseteq \mathscr{N}$ and for every $x \in U_\delta \setminus \{0\}$, $\rho_t(x) \in \mathscr{U}_\varepsilon$ for $t \in [0, T(x))$.

When $\mathscr{D} = \mathscr{N} = \mathbb{R}^n$, we have global finite-time convergence.

Theorem I: For a continuously differentiable function $V : \mathscr{D} \mapsto \mathbb{R}$, such that $k > 0, \alpha \in (0,1)$, where $\alpha$ and $k \in \mathbb{R}$, if there exists a neighborhood of the origin $\mathscr{U} \subset \mathscr{D}$ such that $V$ is positive definite, $\dot{V}$ is negative definite and $\dot{V} + kV^\alpha$ is negative semi-definite on $\mathscr{U}$, where $\dot{V}(x) = \dfrac{\partial V}{\partial x}(x)f(x)$, then the origin of (1) is finite-time stable. Also, the settling time, $T(x)$, is defined as $T(x) = \dfrac{1}{k(1-\alpha)}V(x)^{1-\alpha}$

## 3. Continuous Finite Time Stabilization

Our goal is to find a continuous feedback law, $u = \psi(x,y)$ such that the double integrator defined as,

$$\dot{x} = y, \quad \dot{y} = u \quad (2)$$

is finite-time stabilized.

> **Proposition I**
>
> The origin of the double integrator is globally finite-time stable [Bhat et. al., '98, §III] under the feedback control law $u$ where
>
> $$\psi(x,y) = -\text{sign}(y)|y|^\alpha - \text{sign}(\phi_\alpha(x,y))|(\phi_\alpha(x,y))|^{\frac{\alpha}{2-\alpha}} \quad (3)$$
>
> where $\phi_\alpha(x,y) \triangleq x + \dfrac{1}{2-\alpha}\text{sign}(y)|y|^{\frac{\alpha}{2-\alpha}}$
>
> See Appendix for proof.

**Remarks**: The vector field obtained by using the feedback control law $u$ is locally Lipschitz everywhere except the $x$-axis (denoted $\Gamma$), and the zero-level set $\mathscr{S} = \{(x,y) : \phi_\alpha(x,y) = 0\}$ of the function $\phi_\alpha$. The closed-loop vector field $f_\alpha$ is transversal to $\Gamma$ at every point in $\Gamma \setminus \{0,0\}$

- Every initial condition in $\Gamma \setminus \{0,0\}$ has a unique solution in forward time

- The set $\mathscr{S}$ is positively invariant for the closed-loop system

- On the set $\mathscr{S}$ the closed-loop system is
$$\dot{x} = -\text{sign}(x)\left[(2-\alpha)|x|\right]^{\frac{1}{2-\alpha}} \quad (4)$$
$$\dot{y} = -\text{sign}(y)|y|^\alpha \quad (5)$$

The resulting closed loop system (5) is locally Lipschitz everywhere except the origin and therefore possesses unique solutions in forward time for initial conditions in $\mathscr{S} \setminus \{0,0\}$.

**Example 1**: By choosing $\alpha = \frac{2}{3}$ in (3), we have the phase portrait shown in Figure 1 for the resulting feedback law

$$\psi(x,y) = -y^{\frac{2}{3}} - \left(x + \frac{3}{4}y^{\frac{4}{3}}\right)^{\frac{1}{2}} \quad (6)$$

All trajectories converge to the set $\mathscr{S} = \{(x,y) : x + \frac{3}{4}y^{\frac{4}{3}} = 0\}$ in finite-time. The term $-y^{\frac{2}{3}}$ in (6) makes the set $\mathscr{S}$ positively invariant while the other term $-\left(x + \frac{3}{4}y^{\frac{4}{3}}\right)^{\frac{1}{2}}$ drives the states to $\mathscr{S}$ in finite-time. Therefore, (3) represents an example of a terminal sliding mode control without using discontinuous or high gain feedback.

## 4. Bounded, Continuous Finite-Time Controllers

In the previous section, the designed feedback controller is unbounded, meaning the controller will lead to the "unwinding" phenomenon. Suppose we consider the initial configuration of $(4\pi, 0)$ for the double integrator which coincides with the desired final configuration, we would not need a further control action. Unwinding, however, takes the state $(x,y)$ from $(4\pi, 0)$ to $(0,0)$ making the rigid body rotate twice before reaching a position of rest.

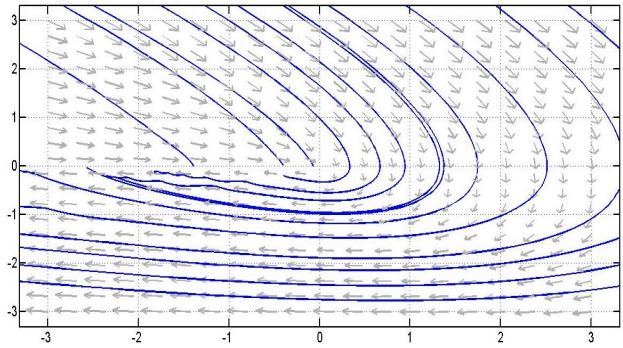

**Figure 1. Double integrator with unsaturated controller** (3). $\alpha = \frac{2}{3}$

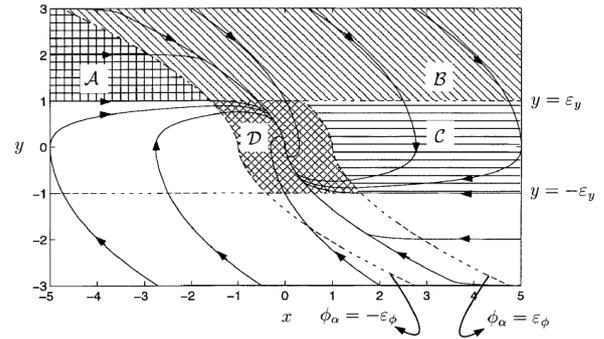

**Figure 2. Double integrator with saturated controller** (8). $\alpha = \frac{1}{3}$

In a spacecraft application, for example, such unwinding can lead to the mismanagement of fuel and and momentum-consuming devices. In order to finite-time stabilize the controller, we saturate its components and define a positive number $\varepsilon$, such that

$$\text{sat}_\varepsilon(y) = y, \qquad |y| < \varepsilon \tag{7}$$

$$= \varepsilon \, \text{sign}(y), \qquad |y| \geq \varepsilon \tag{8}$$

such that $|\text{sat}_\varepsilon(y)| \leq \varepsilon$ for all $y \in \mathbb{R}$

The relation in (8) ensures that the controller does not operate in the nonlinear region by bounding its operating range within the defined perimeter of operation (7). For the double integrator under the feedback control law

$$\psi_{sat}(x,) = -sat_1(y^{\frac{1}{3}}) - sat_1\{(x+\frac{3}{5}y^{\frac{5}{3}})\}^{\frac{1}{5}} \tag{9}$$

obtained from (3) with $\alpha = \frac{1}{3}$ and $\varepsilon = 1$, we see that all trajectories converge to the set $\mathscr{S} = \{(x,y): x+\frac{3}{5}y^{\frac{5}{3}} = 0\}$. But in some phase plane regions, $\psi_{sat}(x,) = 0$.

## 5. The Rotational Double Integrator

Let us denote the motion of a rigid body rotating about a fixed axis with unit moment of inertia as

$$\ddot{\theta}(t) = u(t) \tag{10}$$

where $\theta$ is the angular displacement from some setpoint and $u$ is the applied control. We can rewrite the equation as a first-order equation with $\dot{x} = \theta$ and $\dot{y} = u$. If we require the angular position to be finite-time stable, then the feedback law given in (3) can only finite-time stabilize the origin such that if applied to the rotational double integrator, it leads to the *unwinding* phenomenon.

Therefore, feedback controllers designed for the translational double integrator do not suffice for the rotational double integrator. This disadvantage can be overcome by modifying (3) such that it is periodic in $x$ with period $2\pi$, i.e.,

$$\psi_{rot}(x,y) = -\text{sign}(y)|y|^\alpha$$
$$- \text{sign}(\sin(\phi_\alpha(x,y)))\left|\sin(\phi_\alpha(x,y))\right|^{\frac{\alpha}{2-\alpha}} \tag{11}$$

where $\phi_\alpha$ is same as defined in Proposition 1. For $\alpha = \frac{1}{3}$ and $u = \psi_{rot}$ the resulting phase portrait is shown below

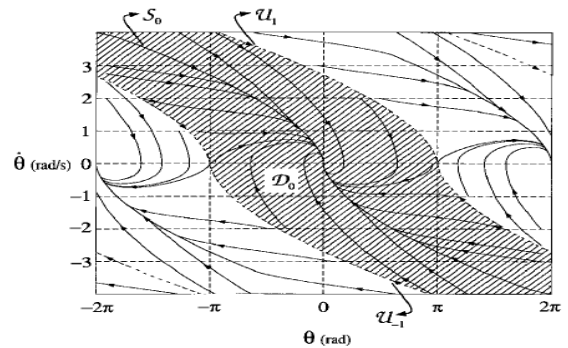

**Figure 3. Rotational double integrator with controller** (11)

**Remarks**

- The closed loop system has equilibrium points at $s_n = (2n\pi, 0)$, $u_n = ((2n+1)\pi, 0)$, $n = \cdots, -1, 0, 1, \cdots$. The equilibrium points $s_n$ are locally finite-time stable in forward time, while the

- points $u_n$ are finite-time saddles. The domain of attraction of the equilibrium point $s$ is $\mathscr{D}_n = \{(x,y) : (2n-1)\pi < \phi_\alpha(x,y) < (2n+1)\pi\}$.

- The shaded region of the plot shows a portion of $\mathscr{D}_0$. The sets $\mathscr{U}_{n-1}$ and $\mathscr{U}_n$ represent the stable manifolds of the equilibrium points $u_{n-1}$ and $u_n$ respectively where $\mathscr{U}_n = \{(x,y) : \phi_\alpha(x,y) = (2n+1)\pi\}$ and $n = \cdots, -1, 0, 1, \cdots$.

- All trajectories starting in the set $\mathscr{D}_n$ converge to the set $\mathscr{S}_n = \{(x,y) : \phi_\alpha(x,y) = 2n\pi\}$ in finite, forward time and to the set $\mathscr{U}_{n-1} \cup \mathscr{U}_n$ in finite, reverse time. The sets $\mathscr{S}_n$ are positively invariant while the sets $\mathscr{U}_n$ are negatively invariant

- The solutions in the figure have no uniqueness to initial conditions lying in any of the sets $\mathscr{U}_n, n = \cdots, -1, 0, 1, \cdots$

- All solutions initialized in $u_n$ are equivalent to the rigid body resting in an unstable configuration and then starting to move spontaneously clockwise or counterclockwise

- Departure from the unstable equilibrium is a unique feature to non-Lipschitzian systems as Lipschitzian systems do not possess solutions that depart from equilibrium

- The desired final configuration is however not globally finite-time stable due to the presence of the unstable equilibrium configuration at $\theta = \pi$. These are saddle points $u_n, n = \cdots, -1, 0, 1, \cdots$

- This is a basic drawback to every continuous feedback controller that stabilizes the rotational double integrator without generating the unwinding effect.

- The desired final configuration in the phase plane corresponds to multiple equilibria in the phase plane meaning every controller that stabilizes the desired configuration stabilizes each equilibria

- But stability, continuous dependence on initial conditions and solutions' uniqueness imply that the domain of attraction of any two equilibrium points in the plane are non-empty, open and disjoint.

- We cannot write $\mathbb{R}^2$ as the union of a collection of disjoint sets.

- Thus, there are initial conditions in the plane that do not converge to the equilibria of the desired final configuration.

- With respect to (11), these initial conditions are the stable manifolds of the unstable configuration.

- The designed controller is practically globally stable as its non-Lipschitzian property increases the sensitivity of the unstable configuration to perturbations

# Appendix: Proof of Theorem I

If we denote $\phi_\alpha(x,y)$ by $\phi_\alpha$ and fix $\alpha \in (0,1)$, we could choose the $\mathscr{C}^2$ Lyapunov function candidate,

$$V(x,y) = \frac{2-\alpha}{3-\alpha}|\phi_\alpha|^{\frac{3-\alpha}{2-\alpha}} + sy\phi_\alpha + \frac{r}{3-\alpha}|y|^{3-\alpha} \quad (12)$$

where $r$ and $s$ are positive numbers. Along the closed loop trajectories,

$$\dot{V}(x,y) = s\phi_\alpha \dot{y} + r|y|^{2-\alpha}\dot{y} + sy\dot{\phi}_\alpha + |\phi_\alpha|^{\frac{1}{2-\alpha}}\dot{\phi}_\alpha$$

$$= s\phi_\alpha \left[-\text{sign}(y)|y|^\alpha - \text{sign}(\phi_\alpha)\right. $$
$$\left. |\phi_\alpha|^{\frac{\alpha}{2-\alpha}}\right] + r|y|^{2-\alpha}\left[-\text{sign}(y)|y|^\alpha - \text{sign}\right.$$
$$\left. (\phi_\alpha)|\phi_\alpha|^{\frac{\alpha}{2-\alpha}}\right] + sy\dot{\phi}_\alpha + |\phi_\alpha|^{\frac{1}{2-\alpha}}\dot{\phi}_\alpha$$
$$(13)$$

But,

$$\dot{\phi}_\alpha = \dot{x} + \dot{y}\,\text{sign}(y)\,|y|^{1-\alpha} \quad (14)$$

From (3), it therefore follows that,

$$\dot{\phi}_\alpha = -\text{sign}(y)\,\text{sign}(\phi_\alpha)|y|^{1-\alpha}|\phi_\alpha|^{\frac{\alpha}{2-\alpha}} \quad (15)$$

Putting (14) into (13), and noting that

$$sy\dot{\phi}_\alpha = -s\,\text{sign}(y\phi_\alpha)|y|^{1-\alpha}|\phi_\alpha|^{\frac{1+\alpha}{2-\alpha}} \quad (16)$$

and $\dot{\phi}_\alpha|\phi_\alpha|^{\frac{1}{2-\alpha}} = -\text{sign}(y)\,\text{sign}(\phi_\alpha)\,|y|^{1-\alpha}|\phi_\alpha|^{\frac{1+\alpha}{2-\alpha}}$
$$(17)$$

we find that,
$$\dot{V}(x,y) = -ry^2 - s|\phi_\alpha|^{\frac{2}{2-\alpha}} - |y|^{1-\alpha}|\phi_\alpha|^{\frac{1+\alpha}{2-\alpha}}$$
$$- s\phi_\alpha \text{sign}(y)|y|^\alpha \quad (18)$$
$$- (r+s)\text{sign}(y\phi_\alpha)|y|^{2-\alpha}|\phi_\alpha|^{\frac{\alpha}{2-\alpha}}$$

**Remarks**

- The obtained Lyapunov derivative in (18) is continuous everywhere since $\alpha \in (0,1)$ and for $k > 0$ and $(x,y) \in \mathbb{R}^2$ the following holds

- If we introduce $x = k^{2-\alpha}, y = ky$ such that

$$\phi_\alpha(k^{2-\alpha}x, ky) = k^{2-\alpha}x \\ - \frac{1}{2-\alpha}\text{sign}(ky)|ky|^{2-\alpha} \\ = k^{2-\alpha}\phi_\alpha(x,y)) \quad (19)$$

- and

$$V(k^{2-\alpha}x, ky) = \frac{2-\alpha}{3-\alpha}\left|\phi_\alpha(k^{2-\alpha}x, ky)\right|^{\frac{3-\alpha}{2-\alpha}} \\ + sky\phi_\alpha(k^{2-\alpha}x, ky) \\ + \frac{r}{3-\alpha}|ky|^{3-\alpha} \quad (20)$$

- such that

$$V(k^{2-\alpha}x, ky) = k^{3-\alpha}V(x,y) \quad (21)$$

- Following a similar logic as in (20), we find that

$$\dot{V}(k^{2-\alpha}x, ky) = k^2\dot{V}(x,y) \quad (22)$$

The results of the previous section imply that for $r > 1$ and $s < 1$, both $V$ and $\dot{V}$ are positive on the set $\mathscr{O} = \{(x,y) : max_{(x,y)\neq(0,0)}|\phi_\alpha|^{\frac{1}{2-\alpha}}, |y| = 1\}$ which is a closed curve encircling the origin.

For every $(x,y) \in \mathbb{R}^2 \setminus \{0,0\}$ there exists $k > 0$ such that $k^{2-\alpha}x, ky \in \mathscr{O}$, the homogeneity properties of (20) and (22) imply a positive definite V and negative semi-definite $\dot{V}$.

From (20), V is radially unbounded so that the set $\mathscr{V} = \{(x,y) : V(x,y) = 1\}$ is compact. Therefore $\dot{V}$ achieves its maximum on the compact set $\mathscr{V}$. If we define $c = -max_{\{(x,y)\in\mathscr{V}\}}\dot{V}(x,y)$, then $\dot{V}(x,y) \leq -c\{V(x,y)\}^{\frac{2}{3-\alpha}}$ for all $(x,y) \in \mathbb{R}^2$ [Bhat et. al. '96]

The homogeneity of (20) and (22) ensures $\dot{V}(x,y) \leq -cV(x,y)^{\frac{2}{3-\alpha}}$ for all $(x,y) \in \mathbb{R}^2$ [Bhat et. al. '96]. Since $\alpha \in (0,1)$ is $\equiv \frac{2}{3-\alpha} \in (0,1)$, we can conclude finite time stability